# Energy transmission control for a Grid-connected modern power system Non-Linear loads with a Series Multi-Stage Transformer Voltage Reinjection with controlled converters

Appalabathula Venkatesh[1], Shankar Nalinakshan[2], S S Kiran[3], Pradeepa H [4]

[1]*Research Scholar, Electrical and Electronics Engineering, The National Institute of Engineering, Mysuru, Karnataka, India*

[2]*Assistant Professor, Electrical and Electronics Engineering, The National Institute of Engineering, Mysuru, Karnataka, India*

[3]*Assistant Professor, Electronics and Communication Engineering, Lendi Institute of Engineering and Technology, Vizianagaram, India*

[4]*Associate Professor, Electrical and Electronics Engineering, The National Institute of Engineering, Mysuru, Karnataka, India*

[1]venkatesh@nie.ac.in, [2] shankar.nalinakshan@nie.ac.in, [3]sskiran88k@gmail.com, [4]pradeep3080@nie.ac.in

***Abstract*** — *The effective way of energy transmission plays a key factor in improving the overall transmission systems efficiency. Many methods are proposed to control the reactive power flow, voltage fluctuations and power factor improvement, The proposed converter topology gives a much significant improvement in transmission systems performance which includes multistage transformers control with the controlled converters along with the series active filters. The overall control strategy which involves the Multistage Voltage Re-Injection Transformer Controlled Converters (MSVRITCC) to reinject the voltages into the grid to compensate the voltages and remaining parameters and power flow control. The proposed topology improves the grid security, flexibility in reaching the desired load requirements with grid adaptability and reduces THD values into a significant values and made the control of power conditioning circuit flexible and easy to perform the voltage compensations in grid to load connected applications. The binary control is used to trigger the power converter circuits which made the controlling much simpler.*

## I. INTRODUCTION

The adequate growth in renewable energy sector in the grid connected power systems has leads to more dependency on distribution systems management in its efficient energy transmission control. The non-linear loads (in current world mainly the sudden raise in the electric vehicles) which are connected to the grid which introduces the harmonics into the system and which intern results into the instable grid connected system [1]-[2]. There are many proposed control strategies are found and implemented to control the parameters presented in the Grid2Load (G2L) connected transmission system [3]-[4]. The proposed control strategy involves the multistage transformer set with the controlled converters and series active filters (SAF). Here, The proposed topology involves the development of Multistage Voltage Re-Injection Transformer controlled converters (MSVRITCC). Similarly, for any type of demands we can implement the desired stage based upon the non-linear load demands [5].

Usually multi level topologies requires DC link capacitors, more number of stages involves more number of DC link capacitors which leads to more chance in getting the failure of those DC link capacitors, proposed topology mainly concentrates on this point and removes the usage of those capacitors and increases the reliability on controlled power converters[6].

One of the solution in removing those capacitor is by using the multi stage cascaded transformers. In this proposed topology single DC link capacitor is used and also reduces the number of power semiconducting devices usage which makes the control and maintenance in a simpler manner. The simulation results shows that the reduction in THD and efficient uni-directional active & reactive power flow control from G2L and voltage fluctuations over conventional multilevel converters[7].

## II. MODELLING OF PROPOSED MSVRITCC

The proposed and implemented block diagram representation of MSVRITCC for the G2L modern power system application is as shown in the Fig. 1.

### A. Mathematical Modelling

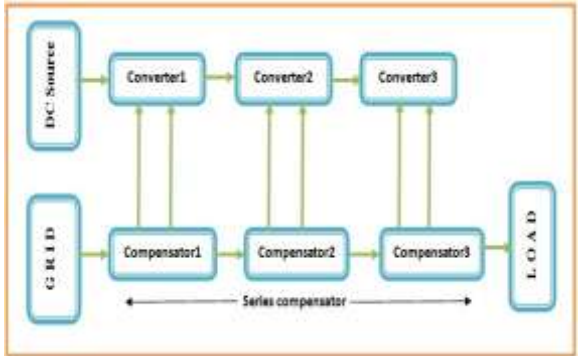

**Fig. 1** Block diagram of 3-stage transformer controlled converters for G2L application

Multi level converter topologies with transformers shows better performance in reducing THD over multi level converters without transformers. The topology is discussed in reference [8] which helpful for modelling and for the mathematical modelling of proposed multistage voltage reinjected (Here,1-stage and 3-stage is considered and designed for analysis) transformer controlled converter (MSVRITCC) for the G2L applications, the following mathematical expressions are needful. The simulation diagram for the proposed 3-stage system is as shown in the Fig. 2

Let,

$v_G \to$ Grid voltage

$v_l \to$ Load voltage

$\dfrac{v_{dc}}{2} \to$ DC link voltage

$v_{Gi}, v_{li} \to$ Corresponding phase's Grid & Load voltage

$v_{mc} \to$ Multistage transformer-controlled converters output voltage

$i \to$ Corresponding phases (R,Y,B)

$p \to$ Corresponding stage (1,2,3…p) (p=3 for 3 stage)

$n_{ip} \to$ Corresponding state of the controlled converters controlled devices

($n_{ip}$ = ON state, $n_{ip}$ = OFF state)

$N_i = 2^{(p-1)}$ Turns ratio of ith phase transformer

Then the controlled converters output each pole voltage can be expressed as

$$v_{mc} = (2n_{ip} - 1)\dfrac{v_{dc}}{2} \qquad (1)$$

The overall proposed MSVRITCC in a modern power systems each phase voltage can be expressed as

$$v_{mc}^* = \sum_{i=R,Y,B} (N_i * v_{pio_i}) \qquad (2)$$

Hence the G2L voltage of the ith corresponding phase can be expressed as follows

$$v_{Gi}, v_{li} = v_{Gli} = v_{mc}^* - \sum_{i=R,Y,B}(N_i * v_{po}) + v_{li}$$

$$v_{Gli} = \sum_{i=R,Y,B}(N_i * v_{pio_i}) - \sum_{i=R,Y,B}(N_i * v_{po}) + v_{li} \qquad (3)$$

The following **Error! Reference source not found.** gives a summary of all pole voltage levels at different states

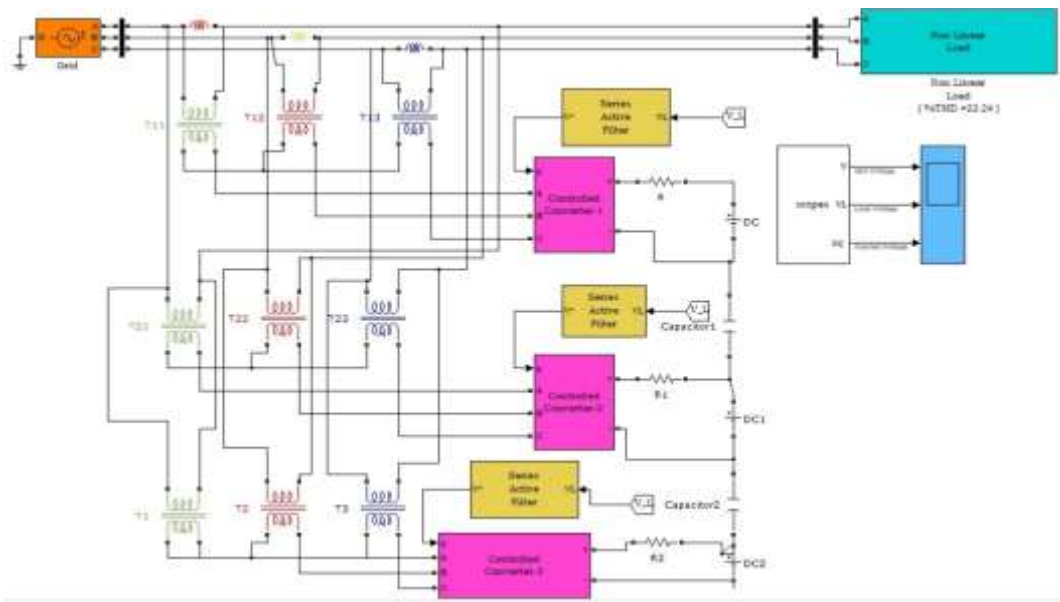

**Fig.2.** Simulation diagram of 3-stage transformer controlled converters for G2L Application

TABLE I
POLE VOLTAGES AT DIFFERENT STATES

| Controlled Device States | State value | Each Phase Output Voltage(Eqn (2)) |
|---|---|---|
| 000 | -7 | $\dfrac{-7v_{dc}}{2}$ |
| 001 | -5 | $\dfrac{-5v_{dc}}{2}$ |
| 010 | -3 | $\dfrac{-3v_{dc}}{2}$ |
| 011 | -1 | $\dfrac{-v_{dc}}{2}$ |
| 100 | 1 | $\dfrac{v_{dc}}{2}$ |
| 101 | 3 | $\dfrac{3v_{dc}}{2}$ |
| 110 | 5 | $\dfrac{5v_{dc}}{2}$ |
| 111 | 7 | $\dfrac{7v_{dc}}{2}$ |

The corresponding simulation results with grid and load voltages for the G2L application along with the proposed MSVRITCC for a 1-stage transformer controlled converter set is as shown in the **Fig.** 3

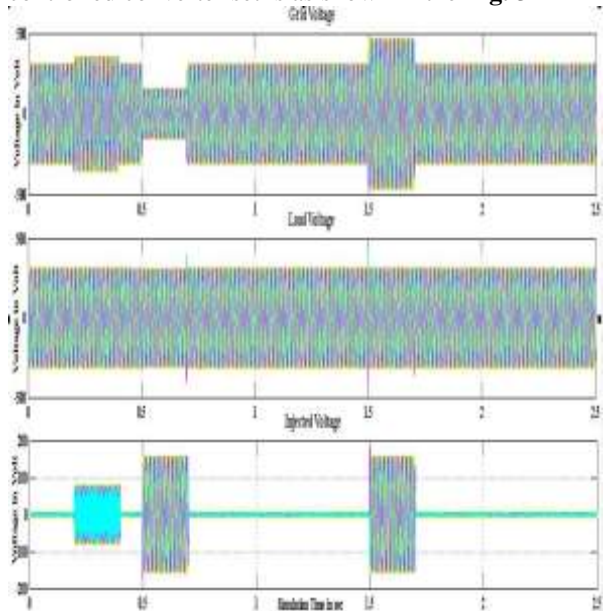

**Fig 3.** Grid, Load & Injected voltages for 1-stage transformer controlled converters for G2L application

Similarly grid and load voltages and the injected voltages for the designed 3-stage transformer controlled converter set is as shown in the **Fig.** 4.

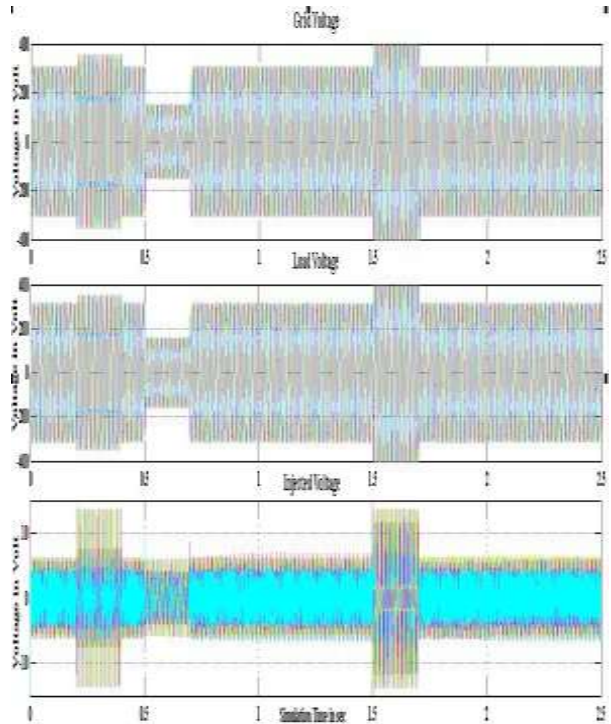

**Fig. 4** Grid, Load & Injected voltages for 3-stage transformer controlled converters for G2L application

The compensated instantaneous active in Watt and reactive powers in Var through the transmitting system supplied by the proposed model is as shown in **Fig. 5**.

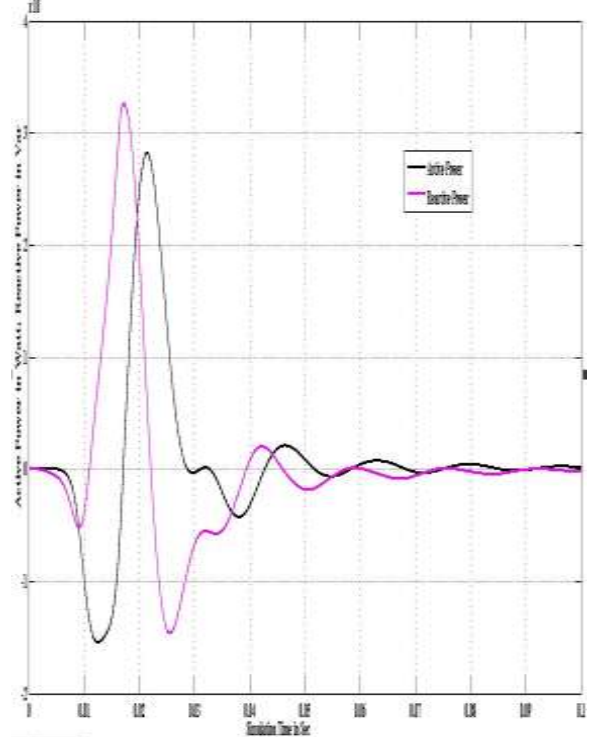

**Fig.5.** Instantaneous Transmitting Powers

The simulation results of proposed system's grid voltage and current are displayed in Fig. 6.

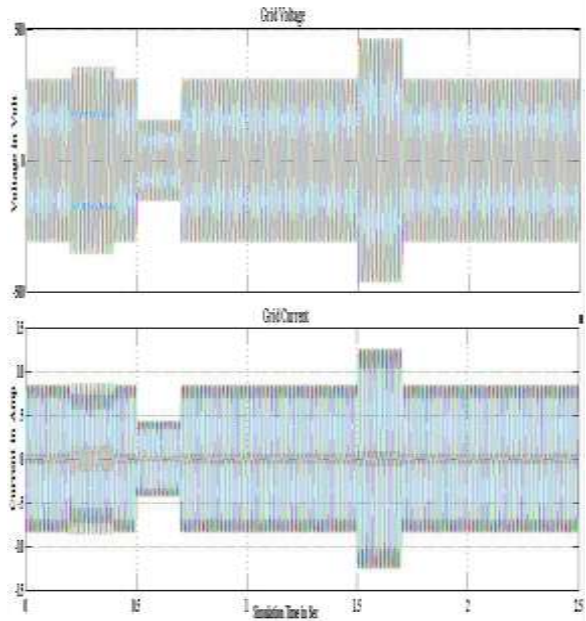

**Fig. 6** Grid voltage and currents

Voltage across the Series Active Filter (SAF) along with the capacitor in phase 'a' is as shown in **Fig. 7.**

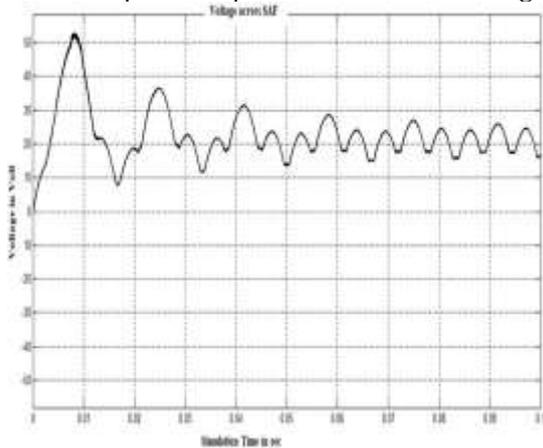

**Fig. 7** Voltage across SAF

The simulation results of transformers secondary side voltages in one phase ('a' phase) is as shown in **Fig. 8**

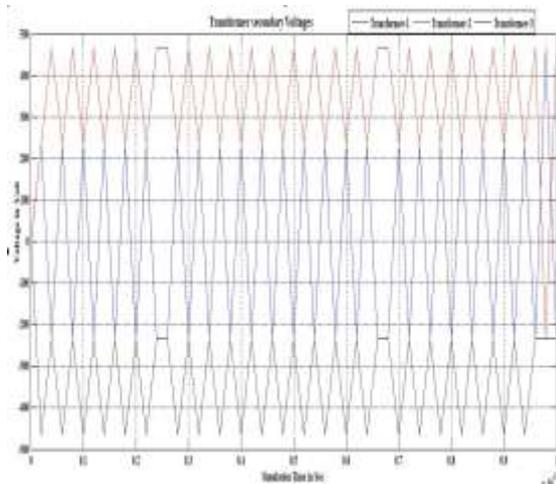

**Fig. 8** Voltage across Secondary's of transformers in 'a' Phase

Source voltage comparison with grid voltage is projected in **Fig. 9**. Which shows that there is a less fluctuations in load which is approximately equal to source voltage with the proposed 3-stage controlled converters.

Estimated THD is 4.83% in the non-linear load voltage with the proposed topology(Considered 50 cycles out of 125 cycles) is shown in the **Fig. 10**.

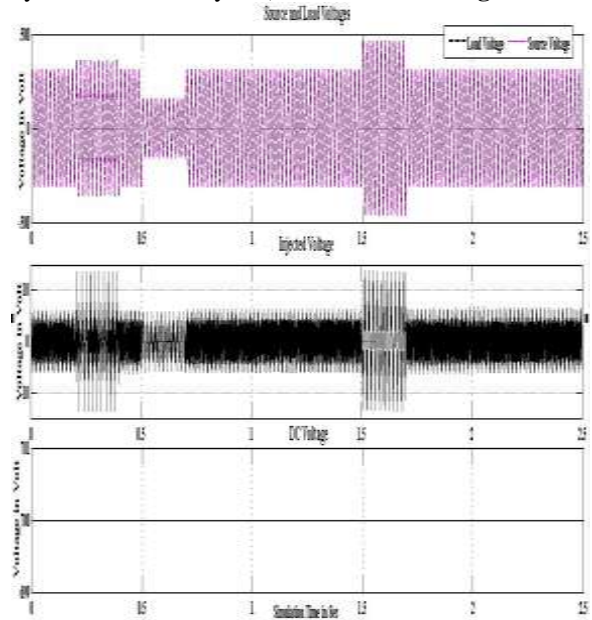

**Fig.9** Comparison of source and load voltages

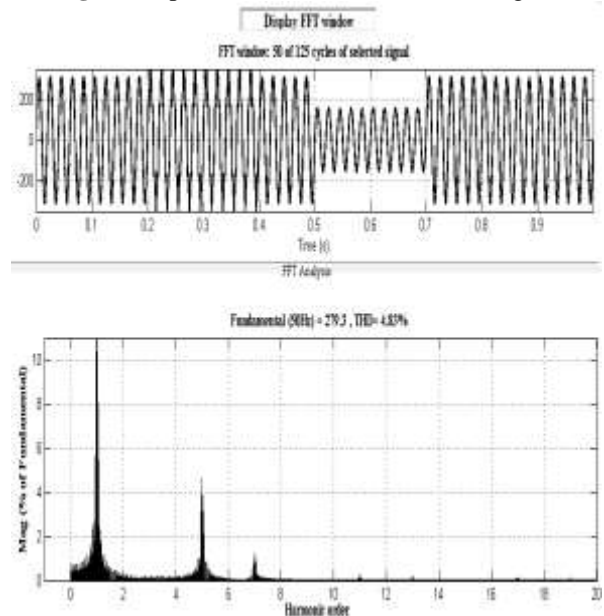

**Fig.10** Estimated THD in Load voltage

. The stability restoration in the load voltage is achieved with the help of auto compensation of voltage levels with the proposed MSVRITCC circuit foe both Swag and Swell of voltages are tested and applied at 0.1 sec. The simulation results at the load end for the Sag(dip in voltage) and Swell(raise in voltage) because of non-linearities in the proposed system are shown in **Fig. 11** and **Fig. 12**.

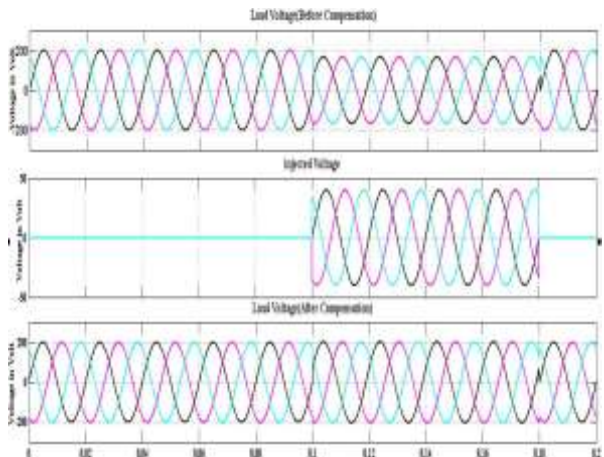

**Fig. 11** Compensation of load voltage due to Sag

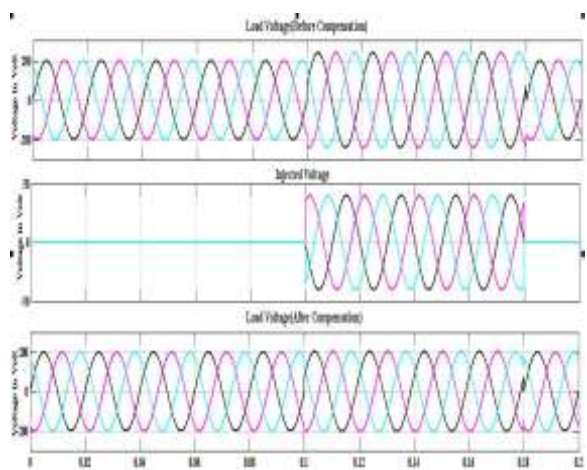

**Fig. 12** Compensation of load voltage due to Swell

The corresponding multi level phase output voltage for the different state values of controlled devices which are listed in Table I and the theoretical calculations are obtained from the Eqn(2). The simulation results of phase 'a' is as shown in **Fig. 13.**

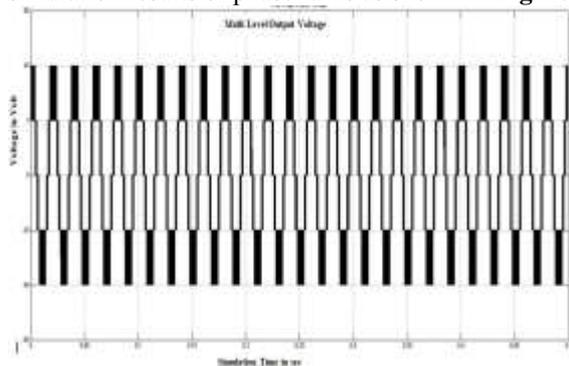

**Fig. 13** Each phase output voltage with different controlled device states(Refer Table I)

### III. APPENDIX

**Transformer Ratings**
Type of Transformer:
2-winding identical transformer(s)
HV side rated voltage: 11kV(rms)
HV winding resistance: 0.002(pu)
HV winding reactance: 0.08(pu)
LV side rated voltage: 3kV(rms)
LV winding resistance: 0.002(pu)
LV winding reactance: 0.08(pu)
Magnetization resistance: 6(pu)
Magnetization reactance: 0.038(pu)
No. of stages: 3

**Control converter parameters:**
Type of converter: IGBT based
Internal ON resistance: $1m\Omega$
Snubber Resistance: $100k\Omega$
Snubber capacitance: Infinite
Load resistance: $60\Omega$
Load reactance: $150\mu F$
Transmitting system per phase inductance: 10mH

### IV. CONCLUSIONS

In this paper, presented a series 1-stage and 3-stage voltage re-injection transformer controlled converters for efficient energy transmission in modern power system. The compensator rating can be increased by increasing the number of stages. The main advantage of reducing number of DC link capacitors with the proposed model (1-capacitor is used) while in other conventional multilevel compensators more DC capacitor links are needed which makes the complications interms of more harmonic distortions. Simulation results are attached for the analysis of power flows and voltage injection by the series transformer controlled converters.